# CORRESPONDENCE BETWEEN PHASOR TRANSFORMS AND FREQUENCY RESPONSE FUNCTION IN RLC CIRCUITS

**Hassan Mohamed Abdelalim Abdalla**

Polytechnic Department of Engineering and Architecture, University of Studies of Udine, 33100 Udine, Italy.

E-mail: mohamedabdelalimabdalla.hassan@spes.uniud.it

**Abstract:**
The analysis of RLC circuits is usually made by considering phasor transforms of sinusoidal signals (characterized by constant amplitude, period and phase) that allow the calculation of the AC steady state of RLC circuits by solving simple algebraic equations. In this paper I try to show that phasor representation of RLC circuits is analogue to consider the frequency response function (commonly designated by FRF) of the total impedance of the circuit. In this way I derive accurate expressions for the resonance and anti-resonance frequencies and their corresponding values of impedances of the parallel and series RLC circuits respectively, notwithstanding the presence of damping effects.

## 1. Introduction and mathematical background

RLC circuits have many applications as oscillator circuits described by a second-order differential equation. The three circuit elements, resistor R, inductor L and capacitor C can be combined in different manners. All three elements in series or in parallel are the simplest and most straightforward to analyze.
RLC circuits are analyzed mathematically in the phasor domain instead of solving differential equations in the time domain.
Generally, a time-dependent sinusoidal function $a(t)$ is expressed in the following way è [1]:
$$a(t) = A_M \sin(\omega t + \alpha) \qquad (1.1)$$
where $A_M$ is the amplitude, $\omega$ is the angular frequency and $\alpha$ is the initial phase of $a(t)$.
The medium value $A_m$ of $a(t)$ is given by:
$$A_m = \frac{1}{T}\int_0^T |a(t)|dt = \frac{2}{\pi}A_M \qquad (1.2)$$
where $T = \frac{2\pi}{\omega}$ is the period of the sinusoidal function $a(t)$.

Usually, a time-dependent sinusoidal function is written also in the following form:
$$a(t) = \sqrt{2}\, A \sin(\omega t + \alpha) \qquad (1.3)$$
where $A$ is the the root mean square of $a(t)$ defined as:
$$A = \sqrt{\frac{1}{T}\int_0^T a^2(t)dt} = \frac{A_M}{\sqrt{2}} \qquad (1.4)$$
The factor $\sqrt{2}$ is called the crest factor.

Considering for instance that we are dealing with invariant-frequency sinusoidal functions, we can represent $a(t)$ by the correspondent phasor $\bar{A}$ via the phasor transform:
$$\bar{A} = \mathcal{S}[a(t)] = \sqrt{2}\, A\, e^{(\omega t + \alpha)} \qquad (1.5)$$
The principal properties of phasor transforms are:
- Linearity:
$$\mathcal{S}[af(t) + bg(t)] = a\mathcal{S}[f(t)] + b\mathcal{S}[g(t)] = a\bar{F} + b\bar{G} \qquad (1.6)$$
- Differentiation:
$$\mathcal{S}[f^{(n)}(t)] = i^n \omega^n \mathcal{S}[f(t)] = i^n \omega^n \bar{F} \qquad (1.7)$$

## 1.1. The Laplace transform

The Laplace transform is one of the most important mathematical tools available for modeling and analyzing the differential equations of linear time-invariant differential equations. It is commonly used in many branches of physics and engineering.

Its main advantage is that differentiation of one coordinate function corresponds to multiplication of the transform by a complex variable $s = \sigma + i\omega$, and thus the differential equation becomes an algebraic equation in $s$. Another advantage of the Laplace transform is that, in solving the differential equations, the initial/boundary conditions are automatically taken care of, and both the particular solution and the complementary solution can be obtained simultaneously.

The Laplace transform of a time domain real function $f(t)$ is defined as [2]
$$F(s) = \mathcal{L}[f(t)] = \int_0^\infty f(t) e^{-st} dt \qquad (1.1.8)$$
And the inverse Laplace transform, given by the Bromwich integral, is
$$f(t) = \mathcal{L}^{-1}[F(s)] = \lim_{T \to \infty} \frac{1}{2\pi i} \int_{\sigma - iT}^{\sigma + iT} F(s) e^{st} ds \qquad (1.1.9)$$
The integral for the inverse Laplace transform is therefore carried out along a generic line parallel to the imaginary axis which belongs to the convergence domain of $F(s)$.

The main properties of the Laplace transforms are:

a- Linearity:
$$\mathcal{L}[af(t) + bg(t)] = a\mathcal{L}[f(t)] + b\mathcal{L}[g(t)] \qquad (1.1.10)$$
b- Differentiation:
$$\mathcal{L}\left[\frac{d^{(n)}f(t)}{dt}\right] = s^n \mathcal{L}[f(t)] - s^{n-1}f(0) - \cdots - f^{(n-1)}(0) = \sum_{k=1}^{n} s^{n-k} \frac{d^{k-1}f(0)}{dt^{k-1}} \qquad (1.1.11)$$
c- Integration:
$$\mathcal{L}\left[\int_0^t f(t) d\tau\right] = \frac{1}{s}\mathcal{L}[f(t)] + \frac{g(0)}{s} \qquad (1.1.12)$$
where $g(0) = \int f(t)dt$ calculated for $t = 0$.

## 1.2. Frequency Response Function and Bode diagrams

The definition of frequency response function (FRF) is based on a characteristic property of linear and time-invariant systems.
We know from systems control theory that transfer functions are mathematical representations to describe the relation between generic input $x(t)$ and output $y(t)$ signals.
If we consider a time-invariant system whose transfer function is the ratio of two polynomial functions and has real-negative poles

$$G(s) = \frac{\mathcal{L}[y(t)]}{\mathcal{L}[x(t)]} = \frac{Y(s)}{X(s)} = \frac{b_m s^m + b_{m-1} s^{m-1} + \cdots + b_1 s + b_0}{a_n s^n + a_{n-1} s^{n-1} + \cdots + a_1 s + a_0} \qquad (1.2.13)$$

the frequency response function $F(\omega)$ is linked to the transfer function $G(s)$ by the following relation:

$$F(\omega) = G(s)|_{s=j\omega} = G(j\omega) \qquad (1.2.14)$$

In general, $G(s)$ can be expressed in the zero/pole/gain form as follows

$$G(s) = K \frac{(s-z_1)(s-z_2)(s-z_3)\ldots(s-z_m)}{(s-p_1)(s-p_2)(s-p_3)\ldots(s-p_n)} \qquad (1.2.15)$$

Zeros and poles can be real (first-order), complex conjugates (second-order) or equal to zero, therefore we can write

$$G(s) = K \frac{s^i (s-z_{i+1})(s-z_{i+2})(s-z_{i+3})(s-z^*_{i+3})}{s^j (s-p_{j+1})(s-p_{j+2})(s-p_{j+3})(s-p^*_{j+3})} \qquad (1.2.16)$$

But first-order zeros and poles of the form $(s-z_k)$ and $(s-p_k)$ can be written in the form $1+\tau_k s$ where $\tau_k$ is the constant time; and second-order zeros and poles of the form $(s-z_k)(s-z^*_k)$ and $(s-p_k)(s-p^*_k)$ can be written in the form $s^2 + 2\zeta_k \omega_{nk} s + \omega^2_{nk}$, where $\omega_{ni}$ is the natural frequency and $\zeta_i$ is the damping factor
Therefore we have

$$G(s) = K \frac{\prod_m (1+\tau_m s) \prod_r (s^2 + 2\zeta_r \omega_{nr} s + \omega^2_{nr})}{s^l \prod_n (1+\tau_n s) \prod_s (s^2 + 2\zeta_s \omega_{si} s + \omega^2_{ns})} \qquad (2.15)$$

where $s^l = s^i / s^j$.
Substituting $s = j\omega$ we obtain the FRF:

$$G(j\omega) = \widetilde{K} \frac{\prod_m (1+\tau_m j\omega) \prod_r (1 + 2j\zeta_r \frac{\omega}{\omega_{nr}} + \frac{\omega^2}{\omega^2_{nr}})}{(j\omega)^l \prod_n (1+\tau_n j\omega) \prod_s (1 + 2j\zeta_s \frac{\omega}{\omega_{ns}} + \frac{\omega^2}{\omega^2_{ns}})} \qquad (1.2.17)$$

where $\widetilde{K} = \frac{K}{\omega^2_{ni}}$.

In this manner we can imagine $G(j\omega)$ as the product of elementary $G_i(j\omega)$, each of them represents a null-pole, a first-order zero/pole or a second-order zero/pole:

$$G(j\omega) = G_1(j\omega) G_2(j\omega) G_3(j\omega) \ldots G_{m+r+l+n+s+1}(j\omega) \qquad (1.2.18)$$

The complex frequency response function can be expressed in terms of real and imaginary parts or in terms of modulus and phase:

$$G(j\omega) = |G(j\omega)| e^{j\emptyset(\omega)} = G_{Re}(j\omega) + j G_{Im}(j\omega) \qquad (1.2.19)$$

where:

- $|G(j\omega)| = \sqrt{G^2_{Re} + G^2_{Im}}$ is the modulus;
- $\emptyset(\omega) = \angle G(j\omega) = \tan^{-1}\left(\frac{G_{Im}}{G_{Re}}\right) + \frac{\pi}{2}[1 - sign(G_{Re})] sign(G_{Re})$ is the phase angle.

$$\text{where: } sign(G_{Re}) = \begin{cases} 1, & G_{Re} > 0 \\ -1, & G_{Re} < 0 \end{cases}$$

Frequency response function FRF is often represented with the so-called *Bode plots*, which depict [3]:
  a) The modulus of the function in terms of a *decibel unit* plotted in a linear scale against frequency in logarithmic scale;
  b) The phase angle of the function in a linear scale against frequency in logarithmic scale.

The modulus of a FRF in *decibel unit* is given by:
$$|G(j\omega)|_{dB} = 20\log_{10}|G(j\omega)| \qquad (1.2.20)$$

Thus (2.16) becomes

$$\begin{aligned}|G(j\omega)|_{dB} &= 20\log_{10}|\widetilde{K}| + \sum_{m} 20\log_{10}|1 + \tau_m j\omega| \\ &+ \sum_{r} 20\log_{10}\left|1 + 2j\zeta_r \frac{\omega}{\omega_{nr}} + \frac{\omega^2}{\omega_{nr}^2}\right| + 20\log_{10}\frac{1}{|(j\omega)^l|} \\ &+ \sum_{n} 20\log_{10}\frac{1}{|1+\tau_n j\omega|} + \sum_{s} 20\log_{10}\frac{1}{|1 + 2j\zeta_s \frac{\omega}{\omega_{ns}} + \frac{\omega^2}{\omega_{ns}^2}|}\end{aligned} \qquad (1.2.21)$$

while the phase angle becomes

$$\begin{aligned}\angle G(j\omega) &= 20\log_{10}|\widetilde{K}| + \sum_{m}\angle|1 + \tau_m j\omega| + \sum_{r}\angle\left|1 + 2j\zeta_r \frac{\omega}{\omega_{nr}} + \frac{\omega^2}{\omega_{nr}^2}\right| \\ &+ \angle\frac{1}{|(j\omega)^l|} + \sum_{n}\angle\frac{1}{|1+\tau_n j\omega|} + \sum_{s}\angle\frac{1}{|1 + 2j\zeta_s \frac{\omega}{\omega_{ns}} + \frac{\omega^2}{\omega_{ns}^2}|}\end{aligned} \qquad (1.2.22)$$

## 2. Discussion

As we know, considering user notation and applying phasor transforms on the passive electrical elements constituting the RLC circuit we find the following properties in the following table [4]:

|  | Governing physical laws | Phasors | $Z = \frac{\overline{U}}{\overline{I}} = R + jX$ | | $Y = \frac{\overline{I}}{\overline{U}} = G + jB$ | |
|---|---|---|---|---|---|---|
|  |  |  | R | X | G | B |
| **Resistor** | $u(t) = Ri(t)$ | $\overline{U} = R\overline{I}$ | R | 0 | $\frac{1}{R}$ | 0 |
| **Capacitor** | $i(t) = C\frac{du(t)}{dt}$ | $\overline{I} = j\omega C\overline{U}$ | 0 | $\frac{-1}{\omega C}$ | 0 | $\omega C$ |
| **Inductor** | $u(t) = L\frac{di(t)}{dt}$ | $\overline{U} = j\omega L\overline{I}$ | 0 | $\omega L$ | 0 | $\frac{-1}{\omega L}$ |

where $Z$ is the impedance, $Y$ is the admittance, $G$ is the conductance and $B$ is the Susceptance of the passive elements of the RLC circuit, $\overline{U}$ and $\overline{I}$ are the voltage and current phasors respectively.
According to the governing physical laws and Kirchhoff principles, the equivalent impedance of the series and parallel RLC circuits are respectively:

$$Z_{series}(s) = Z_R(s) + Z_L(s) + Z_C(s) = R + j\omega L - \frac{j}{\omega C} = R + j\left(\omega L - \frac{1}{\omega C}\right) \qquad (2.23)$$

and

$$Z_{parallel}(s) = \frac{1}{\frac{1}{Z_R(s)} + \frac{1}{Z_L(s)} + \frac{1}{Z_C(s)}} = \frac{1}{\frac{1}{R} + \frac{1}{j\omega L} + j\omega C} \tag{2.24}$$

## 2.1. Impedances of passive electrical elements in Laplace domain

The transfer function given by the ratio between the Laplace transform of the voltage applied to the circuit and the current flowing through the circuit is indeed the electrical impedance:

$$Z(s) = \frac{\mathcal{L}[e_z(t)]}{\mathcal{L}[i_z(t)]} = \frac{E_z(s)}{I_z(s)} \tag{2.1.25}$$

This transfer function is defined assuming that initial conditions are all equal to zero, i.e. $e_z(t=0) = \int e_z t \, dt \,|_{t=0} = 0$.

The impedance transfer function of the passive electrical elements can be derived straightforwardly by taking the Laplace transforms [5] of their constitutive relations:

1. *Impedance of a resistor:*

$$e_R(t) = R i_R(t) \tag{2.1.26}$$

   Applying Laplace transform we get

$$E_R(s) = R I_R(s) \tag{2.1.27}$$

   And therefore

$$Z_R(s) = \frac{E_R(s)}{I_R(s)} = R \tag{2.1.28}$$

2. *Impedance of a capacitor:*

$$e_C(t) = \frac{1}{C} \int i_C(t) dt \tag{2.1.29}$$

   Applying Laplace transform we get

$$E_C(s) = \frac{1}{Cs} I_C(s) \tag{2.1.30}$$

   And therefore

$$Z_C(s) = \frac{E_C(s)}{I_C(s)} = \frac{1}{Cs} \tag{2.1.31}$$

3. *Impedance of an inductor:*

$$e_L(t) = L \int i_L(t) dt \tag{2.1.32}$$

   Applying Laplace transform we get

$$E_L(s) = Ls I_L(s) \tag{2.1.33}$$

   And therefore

$$Z_C(s) = \frac{E_L(s)}{I_L(s)} = Ls \tag{2.1.34}$$

obtaining the same results of the table above.

According to the constitutive equations for the resistor, capacitor and inductor passive elements given above, the equivalent impedances of the series and parallel RLC circuits are respectively:

$$Z_{series}(s) = Z_R(s) + Z_L(s) + Z_C(s) = R + Ls + \frac{1}{Cs} \quad (2.1.35)$$

and

$$Z_{parallel}(s) = \frac{1}{\frac{1}{Z_R(s)} + \frac{1}{Z_L(s)} + \frac{1}{Z_C(s)}} = \frac{1}{\frac{1}{R} + \frac{1}{Ls} + Cs} \quad (2.1.36)$$

Now if we consider the FRF of the last two transfer functions we get readily equations (2.23) and (2.24) respectively.

## 2.2. RLC parallel circuit

The FRF of the impedance transfer function is

$$Z(j\omega) = \frac{1}{\frac{1}{R} + \frac{1}{j\omega L} + j\omega C} = \frac{1}{\frac{Lj\omega + R - RLC\omega^2}{j\omega LR}} = RL\frac{j\omega}{R + Lj\omega - RLC\omega^2}$$

$$= L\frac{j\omega}{1 + 2j\zeta\frac{\omega}{\omega_n} - \frac{\omega^2}{\omega_n^2}} \quad (2.2.37)$$

Therefore we obtain the natural frequency $\omega_n = \frac{1}{\sqrt{LC}}$ and the damping factor

$$\zeta = \frac{L\omega_n}{2R} = \frac{\varepsilon}{\omega_n} \quad (2.2.38)$$

where $\varepsilon = \frac{1}{2RC}$ is the damping attenuation.

Expression (2.2.37) is characterized by constant gain $L$, a null zero and simple complex conjugates poles, therefore the modulus and the phase are respectively thus calculated:

$$|Z(j\omega)|_{dB} = 20\log_{10}|L| + 20\log_{10}\omega - 20\log_{10}\sqrt{\left[1 - \left(\frac{\omega^2}{\omega_n^2}\right)\right]^2 + \left(\frac{4\zeta^2\omega^2}{\omega_n^2}\right)} \quad (2.2.39)$$

$$\angle Z(j\omega) = 90° + \tan^{-1}\left(-\frac{2\zeta\frac{\omega}{\omega_n}}{1 - \left(\frac{\omega}{\omega_n}\right)^2}\right) + \frac{\pi}{2}[1 - sign(Z_{Re})]sign(Z_{Re}) \quad (2.2.40)$$

Let's analyze the second-order term; the resonance is caused principally by the presence of the complex conjugate poles, therefore the maximum of $|Z(j\omega)|_{dB}$ corresponds to the minimum of the function $\left[1 - \left(\frac{\omega^2}{\omega_n^2}\right)\right]^2 + \left(\frac{4\zeta^2\omega^2}{\omega_n^2}\right)$, which gives the value of the resonance frequency $\omega_r = \omega_n\sqrt{1 - 2\zeta^2}$.

Now we wish to calculate the amplitude of $|Z(j\omega)|_{dB}$ when $\omega = \omega_r$, i.e. the maximum magnitude of $|Z(j\omega)|_{dB}$:

$$|Z(j\omega_r)|_{dB} = 20\log_{10}|L| + 20\log_{10}\omega_r - 20\log_{10}\sqrt{\left(2\zeta\sqrt{1-\zeta^2}\right)}$$

$$= 20\log_{10}|L| + 20\log_{10}\frac{1}{\sqrt{LC}}\sqrt{1-\frac{L}{2R^2C}}$$

$$- 20\log_{10}\sqrt{\left(\frac{\sqrt{L}}{R\sqrt{C}}\sqrt{1-\frac{L}{4R^2C}}\right)} \tag{2.2.41}$$

## 2.3. RLC series circuit

The FRF of the impedance transfer function is now

$$Z(j\omega) = R + j\omega L + \frac{1}{j\omega C} = \frac{1 + RLj\omega - LC\omega^2}{j\omega C} = \frac{1}{C}\frac{1 + 2j\zeta\frac{\omega}{\omega_n} - \frac{\omega^2}{\omega_n^2}}{j\omega} \tag{2.3.42}$$

Therefore we obtain the natural frequency $\omega_n = \frac{1}{\sqrt{LC}}$ and the damping factor

$$\zeta = \frac{R}{2L\omega_n} = \frac{\varepsilon}{\omega_n} \tag{2.3.43}$$

where $\varepsilon = \frac{1}{2RL}$ is the damping attenuation.

Expression (2.3.42) is characterized by constant gain $C^{-1}$ a null pole and simple complex conjugates zeros, therefore the modulus and the phase are respectively thus calculated:

$$|Z(j\omega)|_{dB} = 20\log_{10}\left|\frac{1}{C}\right| - 20\log_{10}\omega + 20\log_{10}\sqrt{\left[1-\left(\frac{\omega^2}{\omega_n^2}\right)\right]^2 + \left(\frac{4\zeta^2\omega^2}{\omega_n^2}\right)} \tag{2.3.44}$$

$$\angle Z(j\omega) = -90° + \tan^{-1}\left(-\frac{2\zeta\frac{\omega}{\omega_n}}{1-\left(\frac{\omega}{\omega_n}\right)^2}\right) + \frac{\pi}{2}[1 - sign(Z_{Re})]sign(Z_{Re}) \tag{2.3.44}$$

Let's analyze the second-order term; the anti-resonance is caused principally by the presence of the complex conjugate zeros, therefore the minimum of $|Z(j\omega)|_{dB}$ corresponds to the minimum of the function $\left[1-\left(\frac{\omega^2}{\omega_n^2}\right)\right]^2 + \left(\frac{4\zeta^2\omega^2}{\omega_n^2}\right)$, which gives the value of the anti-resonance frequency $\omega_{ar} = \omega_n\sqrt{1-2\zeta^2}$.

Now we wish to calculate the amplitude of $|Z(j\omega)|_{dB}$ when $\omega = \omega_{ar}$, i.e. the minimum magnitude of $|Z(j\omega)|_{dB}$ :

$$|Z(j\omega_r)|_{dB} = 20\log_{10}\left|\frac{1}{C}\right| - 20\log_{10}\omega_{ar} + 20\log_{10}\sqrt{\left(2\zeta\sqrt{1-\zeta^2}\right)}$$

$$= 20\log_{10}\left|\frac{1}{C}\right| - 20\log_{10}\frac{1}{\sqrt{LC}}\sqrt{1-\frac{CR^2}{2L}}$$

$$+ 20\log_{10}\sqrt{\left(\frac{R\sqrt{C}}{\sqrt{L}}\sqrt{1-\frac{CR^2}{4L}}\right)} \tag{2.3.45}$$

## 3. Conclusion

Almost all circuit analysis textbooks treat RLC circuits without considering damping effects, equating thus the natural angular frequency $\omega_n$ with the resonance/anti-resonance frequency; this is true only in theoretical LC circuits, inasmuch as resistive effects are unavoidable in real circuits even if a resistor is not specifically included as a component.
In this paper I have demonstrated the strong correspondence between phasor transforms and the frequency response function of impedances. In this way, considering Bode plots, we can obtain accurate measurements of resonance/anti-resonance angular frequencies in RLC circuits and the values of their corresponding impedances in dB, notwithstanding the presence of damping effects. This paper is intended for graduate students and for the instructors who teach the theory of AC electric circuits.

## 4. Acknowledgments

Dedicated to Giulio D'Onofrio, hoping things would go better.
I would like to thank Marina Radolovic for linguistic corrections and her constant help.